\documentclass[aps,prd,letterpaper,showpacs,twocolumn,preprintnumbers,floatfix,superscriptaddress]{revtex4-1}
\usepackage{graphicx,dcolumn,bm,epsfig,cancel,hyperref}
\usepackage{amsmath}
\usepackage{amssymb, times}
\newcommand{\comment}[1]{{}}
\usepackage{slashed}
\usepackage{color}

\begin{document}
\title{Electron Flavored Dark Matter}
\author{Wei Chao}
\affiliation{Center for Advanced Quantum Studies, Department of Physics, 
Beijing Normal University, Beijing, 100875, China}
\email{chaowei@bnu.edu.cn}
\author{Huai-Ke Guo}
\affiliation{
CAS Key Laboratory of Theoretical Physics, Institute of Theoretical Physics, \\
Chinese Academy of Sciences, Beijing 100190, China}
\email{ghk@itp.ac.cn}
\author{Hao-Lin Li}
\affiliation{
Amherst Center for Fundamental Interactions, Department of Physics, \\
University of Massachusetts-Amherst, 710 North Pleasant St. Amherst, MA 01003 U.S.A.}
\email{haolinli@physics.umass.edu}
\author{Jing Shu}
\affiliation{
CAS Key Laboratory of Theoretical Physics, Institute of Theoretical Physics, \\
Chinese Academy of Sciences, Beijing 100190, China}
\affiliation{
School of Physical Sciences, University of Chinese Academy of Sciences, Beijing 100190, P. R. China
}
\affiliation{
CAS Center for Excellence in Particle Physics, Beijing 100049, China
}
\email{jshu@itp.ac.cn}


\begin{abstract}
In this paper we investigate the phenomenology of the electron flavored Dirac dark matter with two types of portal interactions. We analyze constraints from the electron magnetic moment anomaly, LHC searches of singly charged scalar, dark matter relic abundance as well as direct and indirect detections. 
Our study shows that the available parameter space is quite constrained, but there are parameter space that is compatible with the current data.
We further show that the DAMPE cosmic ray electron excess, which indicates cosmic ray excess at around 1.5 TeV, can be interpreted as the annihilation of dark matter into electron positron pairs in this model.
\end{abstract}

\maketitle

\section{ Introduction}
About $80\%$ of the mater in our Universe is made of dark matter (DM). Among many different DM scenarios, Weakly Interacting Massive Particle (WIMP) remains to be an interesting candidate since it has a strong connection with physics beyond Standard Model at the TeV scale and can be probed through both direct and indirect detections. There has been several indirect DM searches~\cite{Aguilar:2013qda,FermiLAT:2011ab,Adriani:2017efm} which indicate a possible excess for the electron positron cosmic ray spectrum in the 100 GeV $\sim$ TeV energy region. Most recently the DM Particle Explorer (DAMPE) has reported their first result~\cite{Ambrosi:2017wek}, which observes an excess in the electron positron cosmic ray spectrum up to several TeV. For possible theoretical interpretations, see Refs~\cite{Yuan:2017ysv,Fan:2017sor,Fang:2017tvj,Fang:2017tvj,Duan:2017pkq,Gu:2017gle,Zu:2017dzm,Tang:2017lfb,Huang:2017egk,Athron:2017drj,Cao:2017ydw,Liu:2017rgs,Duan:2017qwj,Gu:2017bdw,Chao:2017yjg} for detail. This experiment has several good features in terms of probing the electron positron cosmic ray spectrum. 1): it has a good energy resolution in the high energy region ($<1.2 \%$ for $E>100$ GeV), therefore can be used to detect the line or sharp structure of the particle spectrum in the future. 2): The large detector can have a high statistics. 3): It measures both the low and high energy electron positron cosmic ray spectrum. The first feature is interesting since it can be used to probe possible line or sharp structure of the particle spectrum, which give us much more information and has important implications on possible dark matter interpretations~\footnote{The current DAMPE data, however, does not favor significantly on a sharp excess, especially considering that the two bins next to the 1.4 TeV bin actually has a large deficit comparing to the smoothly broken power-law.}. 

While there has been various studies of dark matter annihilating or decaying into leptons, here we consider a novel class of dark matter which we call ``electron flavored dark matter". In this case, the dark matter carries the electron number and annihilate into electron pairs through the t-channel mediator which could result a possible sharp structure in the electron positron cosmic ray spectrum. We study various constraints such as the electromagnetic properties of the dark matter, fitting the electron magnetic moment, collider phenomenology, dark matter relic abundance and direct detections. More importantly we systematically investigate constraints from indirect DM searches, such as AMS-02~\cite{Aguilar:2013qda}, IceCube~\cite{Aartsen:2016pfc} and CMB~\cite{Ade:2015xua}. Interestingly, combing all constraints together, we show that the electron-flavored DM can address the DAMPE cosmic ray excess without conflicting with any current constraints. 

The remaining of the paper is organized as follows: In section II we introduce the model in detail. Section III is focused on the constraints of electron magnetic moment and colliders. In section IV and V we study their implication on DM relic abundance as well as direct and indirect DM searches. The last part is the concluding remarks.
 
\section{\label{sec:model}Model}
\begin{figure}[t]
\centering
   \includegraphics[width=0.45\columnwidth]{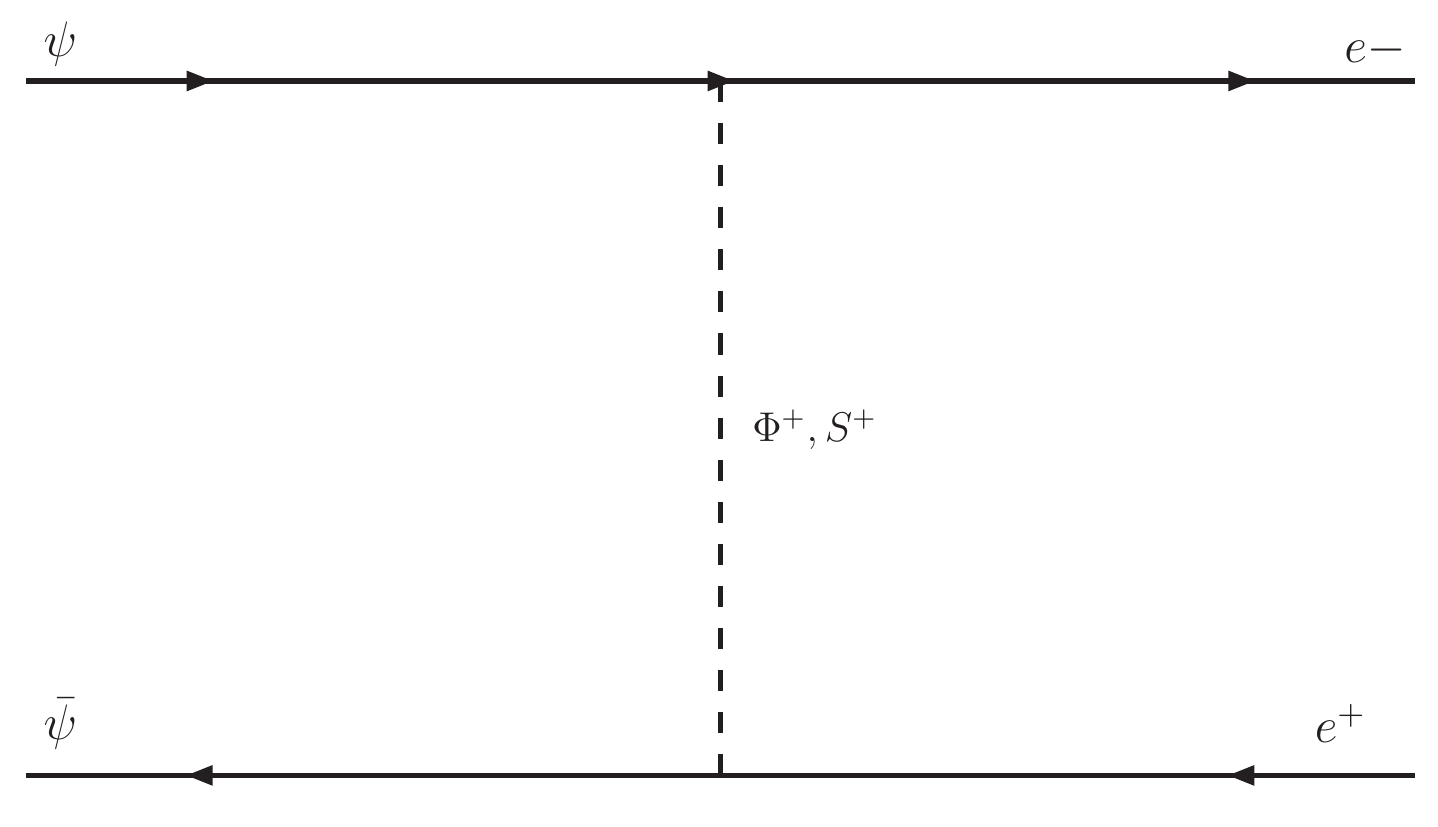}
   \hspace{0.5cm}
   \includegraphics[width=0.45\columnwidth]{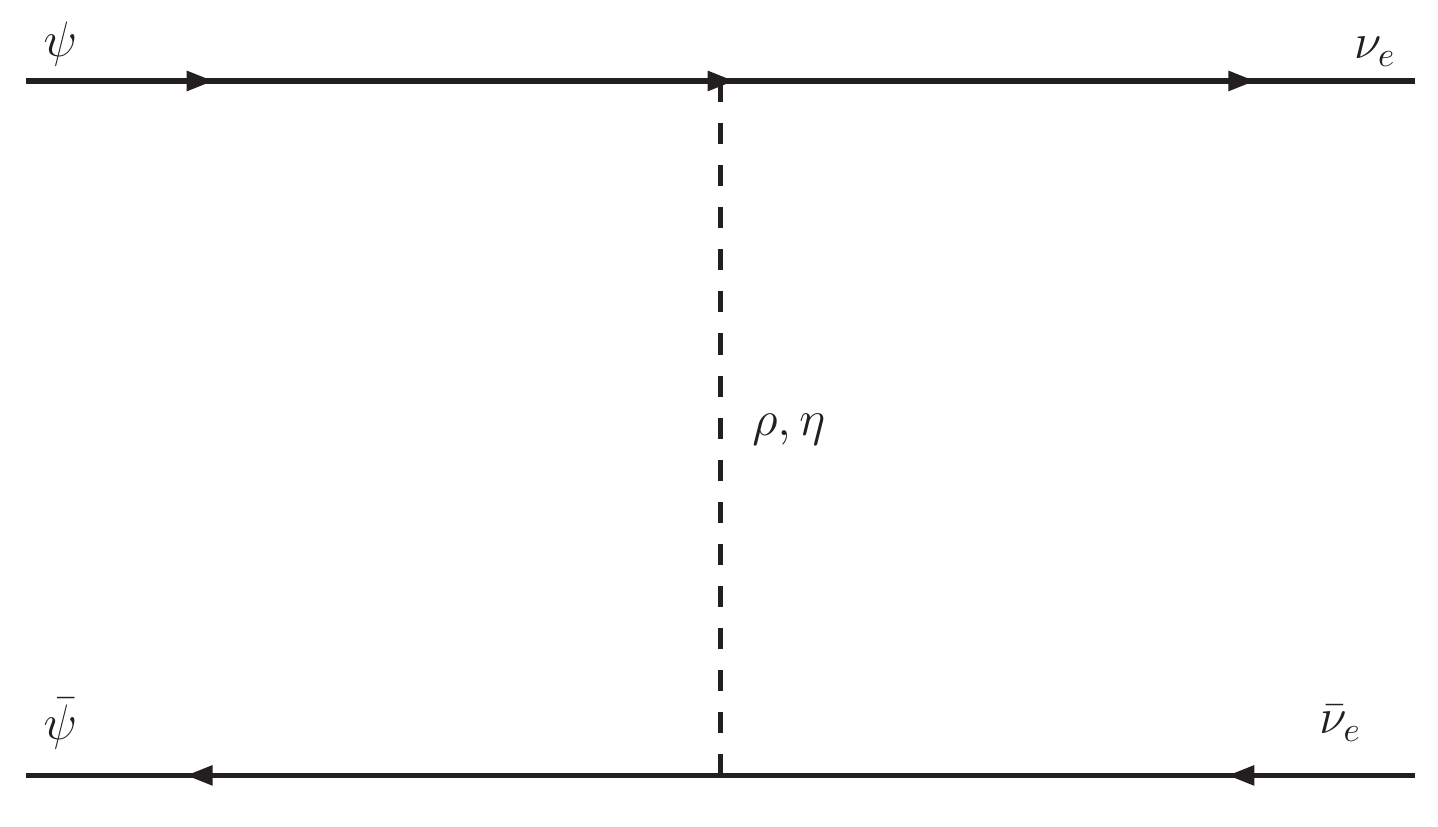}
 \caption{\label{fig:relic} Annihilation channels of electron flavored Dirac DM.}
 \end{figure}

We consider a Dirac DM $\psi$ which couples to the electron via portal interactions of 
two types consistent with EW symmetry. 
For the first type, $\psi$ couples to the first generation  lepton SU(2) doublet $l_L^1$ with the interaction 
\begin{eqnarray}
  \text{Model I}: \quad \quad -  \mathcal{L}_{\text{I}} = \kappa_1^{}  \overline {{\ell^1_L}} \tilde \Phi \psi + {\rm h.c.}  ,
\end{eqnarray}
where $\Phi^T\equiv(\Phi^+, (\rho + i \eta)/\sqrt{2})$ is an inert scalar doublet and 
$\tilde{\Phi} = i \sigma_2 \Phi^{\ast}$. 
The model has a $Z_2$ symmetry in which new particles are odd while all SM particles are even. 
The coupling of $\psi$ with other lepton doublets can be forbidden by another $Z_2$ symmetry in which $\psi$ and $\ell_L^1$ are odd while all other particles are even. 
Note the mass of $\psi$ needs to be smaller than the neutral and charged scalars so as to be stable DM candidate.
For this scenario, the DM can annihilate to $e^+ e^-$ and $\bar{\nu}_{e} \nu_e$ via $t$-channel exchanges of 
charged scalars or neutral scalars shown in Fig.~\ref{fig:relic}.
%
%
%

For the second type model, the DM is coupled to the electroweak singlet $e_R$ via
\begin{eqnarray}
  \text{Model II}: \quad \quad -  \mathcal{L}_{\text{II}} = \kappa_2^{} \overline \psi S^+ {e_R}  + {\rm h.c.}  ,\;
\end{eqnarray}
where $S^+$ is a singly charged scalar singlet. In this case, the DM annihilates only into $e^+ e^-$ through 
exchanges of the charged scalar, corresponding to the left panel of Fig.~\ref{fig:relic}. 
The discrete flavor symmetry is the same as these in Model I.

\begin{figure}[t]
\centering
  \includegraphics[width=0.4\columnwidth]{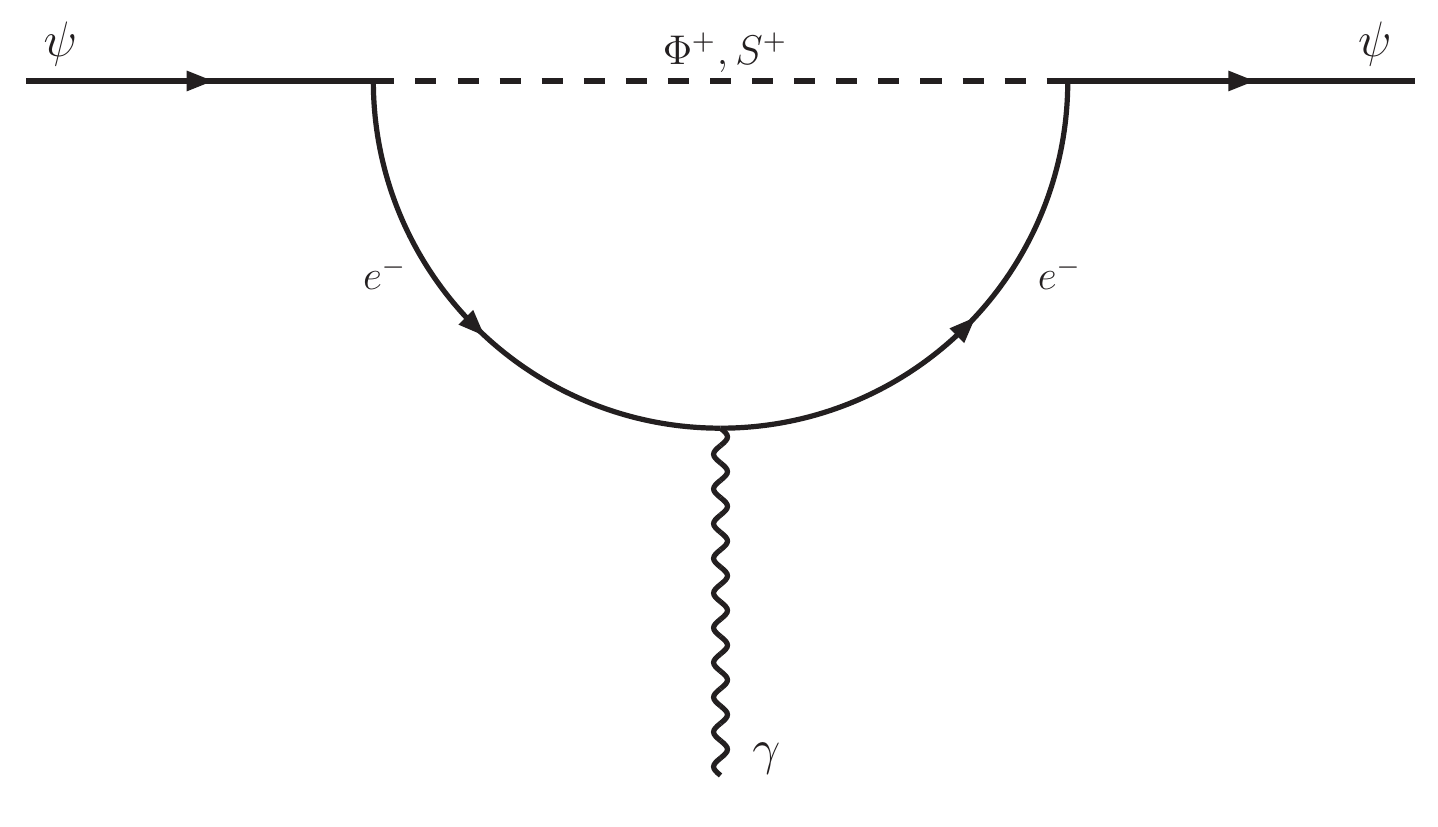}
  \includegraphics[width=0.4\columnwidth]{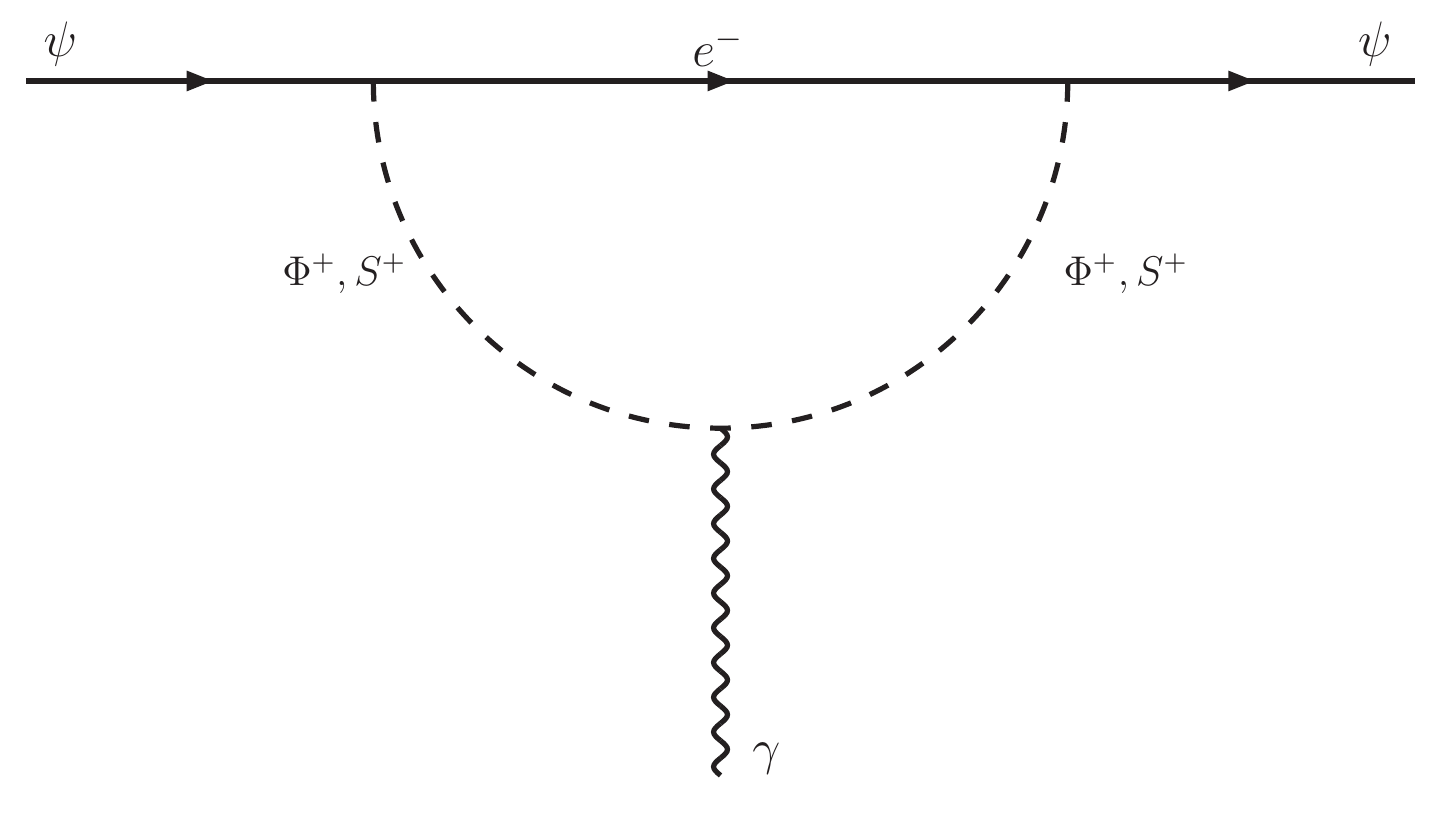}
\caption{\label{ddloopgamma} Feynman diagrams contributing to the DM electromagnetic form factors.}
\end{figure}

For the benefit of the direct detection, one needs to calculate the electromagnetic form factors of the DM, which arise at one loop level from the relevant Feynman diagrams shown in fig.~\ref{ddloopgamma}.
These diagrams generate the following effective DM-photon effective interactions~\cite{Chao:2016lqd}:
\begin{eqnarray}
b_{\psi} \bar{\psi} \gamma^{\mu} \psi \partial^{\nu} F_{\mu\nu} 
+ c_{\psi} \bar{\psi} \gamma^{\mu} \gamma^5 \psi \partial^{\nu} F_{\mu\nu} +
\frac{\mu_{\psi}}{2} \bar{\psi} \sigma^{\mu\nu} \psi F_{\mu\nu} , \ \  \ 
  \label{eq:dm-photon}
\end{eqnarray}
where $b_\psi$ is the DM charge radius, $c_\psi$ is the axial charge radius or anapole moment and 
$\mu_\psi$ is the magnetic moment. For both models, the results can be summarized by a uniform 
set of formulae~\cite{Chao:2016lqd}:
\begin{eqnarray}
  \mu_{\psi} &=& -\frac{e m_{\psi}  \kappa_i^2 }{64\pi^2} \int^1_0 dx \frac{x(1-x) }{\Delta_i} \; , \nonumber \\
  b_{\psi} &=&{e \kappa^2_i \over 32 \pi^2 } \int_0^1 d x \Big\{{ x^3 -2(1-x)^3\over 6 \Delta_i }\nonumber \\
&& \hspace{1cm} + {(x-1)^3 (x^2 m_\psi^2+{m_e^2})  +2(1-x)x^4 m_\psi^2 \over 6 \Delta_i^2 }  \Big\} , \label{electromag} \nonumber \\
  c_{\psi} & =&  \frac{(-1)^{i-1}e \kappa_i^2 }{192 \pi^2} \int^1_0 \frac{dx}{\Delta_i^2} \left\{  (-3x^3+6x^2-6x+2) x m_i^2 \right. \nonumber \\
		  && \hspace{1.5cm} \left.  +(-2x^4+6x^3-9x^2+7x-2) x m_{\psi }^2 \right\} ,\nonumber
\label{eq:dmresults}
\end{eqnarray}
where  $\Delta_i= x {m}_i^2 + x(x-1)m_\psi^2 + (1-x){m_e^2}$ and the index
``$i$'' denotes the charged scalar in each model. 
%
%
Since $m_e$ is much smaller than the momentum transfer$\sqrt{|q^2|}\approx 50\text{MeV}$,
the infrared divergence as $m_e \rightarrow 0$ is cut-off by the momentum transfer~\cite{Bai:2013iqa}. Therefore we 
replace $m_e$ by $50\text{MeV}$ in the numerical calculation.

\section{Constraints}
Similar to other lepton flavored DM scenarios, the electron DM couplings here face constraints from 
precision measurement of the electron magnetic dipole moment(MDM) as well as searches at the colliders.

%
%

\begin{figure*}[t!]
\centering
  \includegraphics[width=0.8\columnwidth]{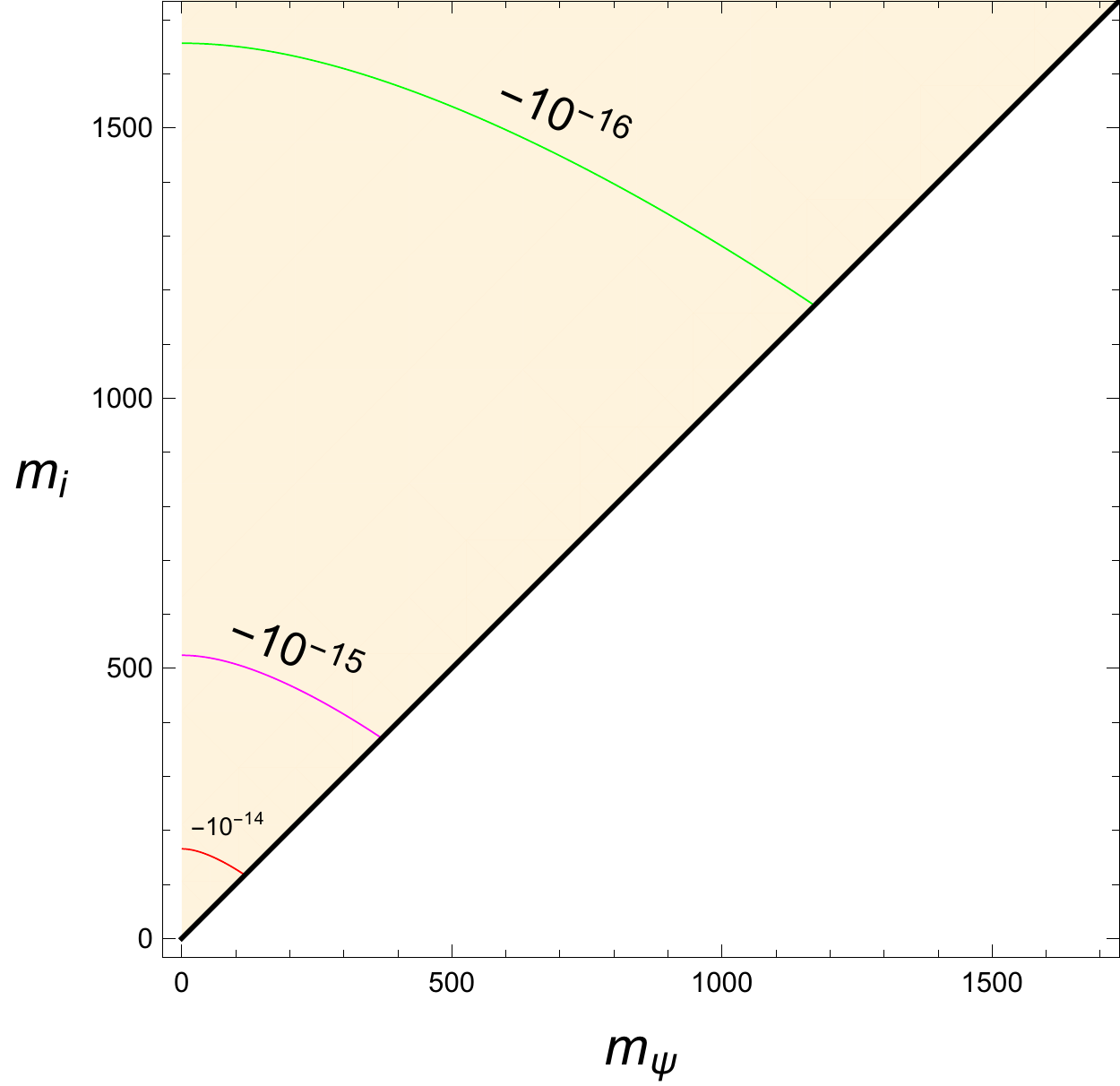}
  \hspace{1cm}
  \includegraphics[width=0.77\columnwidth]{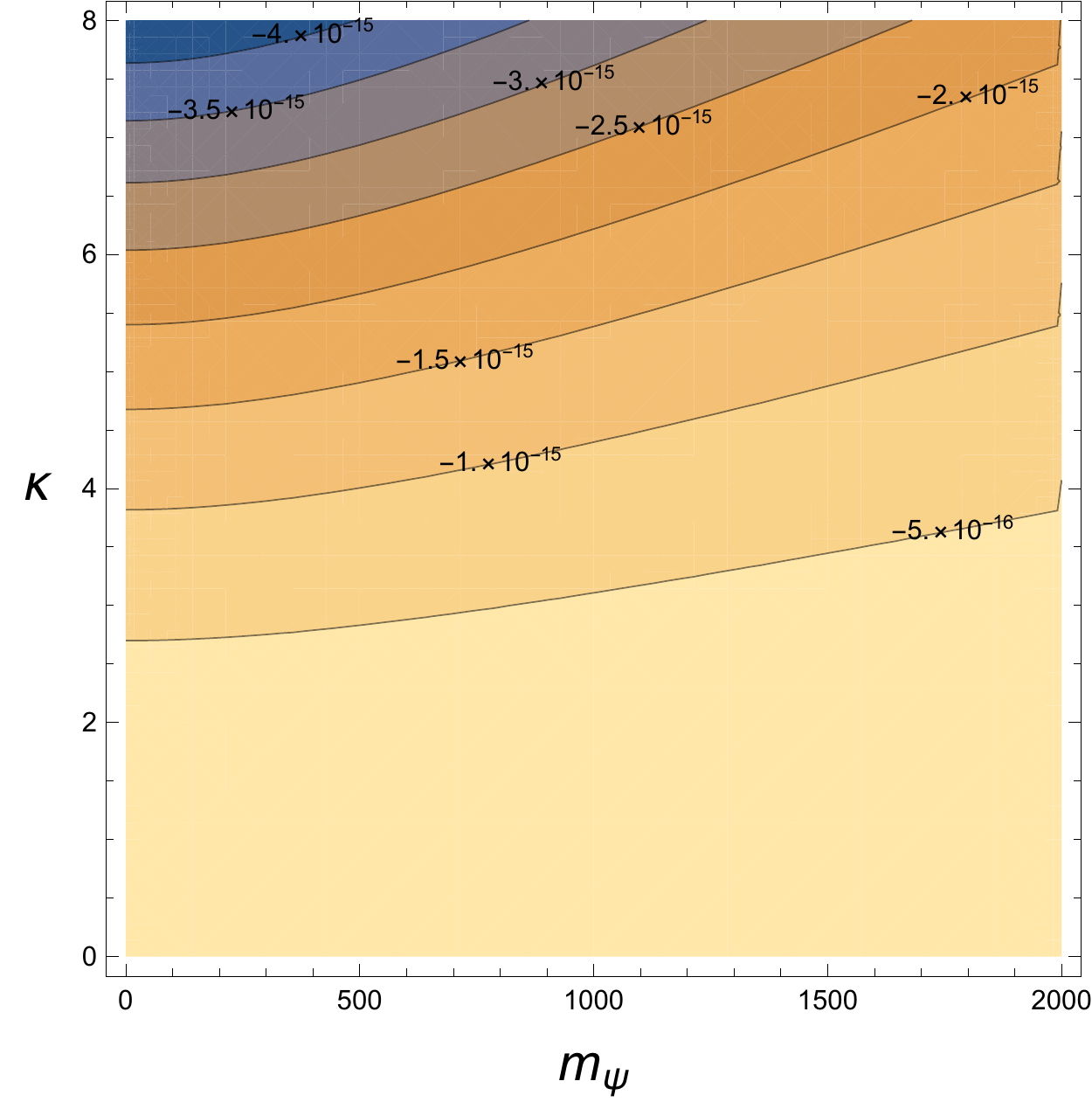}
  \caption{\label{fig:eMDM}The left plot shows eMDM for fixed $\kappa_1=1$ or $\kappa_2=1$ in the plane of DM 
  mass($m_{\psi}$) and charged scalar mass $m_i$. The right plot fixes $m_i=2\text{TeV}$ and shows
  the contours of $\delta a_e$ in the plane of the DM mass and the coupling $\kappa_1$ or $\kappa_2$, 
collectively denoted as $\kappa$ here.
}
\end{figure*}
%
The 2017 PDG tabulates the measured eMDM anomaly as $(1159.65218091\pm 0.00000026)\times 10^{-6}$ where 
the uncertainty is only $10^{-10}$ of
the central value~\cite{Patrignani:2016xqp}, constituting one of the most precisely measured physical constants. 
At one-loop, the modification of eMDM anomaly is
\begin{eqnarray}
  &&  \delta a_e = 
  \frac{Y_i m_e^2}{4\pi^2 m_i^2}
   \frac{6 x_i^2 \ln(x_i) + 6 x_i -1 - 3 x_i^2 - 2 x_i^3}{24 (1-x_i)^4} ,  \  \
\end{eqnarray}
with $x_i \equiv m^2_{\psi}/m_i^2$ where the index ``$i$'' denotes the charged scalar 
$\Phi^+$($Y_{\Phi^+}=\kappa_1^2$) for model I and $S^+$($Y_{S^+}=\kappa_2^2$) for model II.
The $x_i$ dependent factor above increases from $-1/24$ for $x\rightarrow 0$ to $-1/48$ for $x\rightarrow 1$,
leading to a negative $\delta a_e$.
Compared with muon MDM, the eMDM is not so sensitive to the new physics contribution since 
$\delta a_e$ is suppressed by the factor $m_e^2/m_{\mu}^2 \approx 10^{-5}$. 
In Fig.~\ref{fig:eMDM}, the 
contours of eMDM are shown: the left panel shows in the $(m_{\psi}, m_i)$ plane by fixing $\kappa=1$ while 
the right panel shows in the plane $(\kappa, m_{\psi})$ by fixing $m_i=2\text{TeV}$.
As is clear from these
figures, $\delta a_e$ is much smaller than current experimental un-certaintity and eMDM imposes no severe
constraint on the couplings $\kappa_1$ for model I and $\kappa_2$ for model II.

Notice that when both types of portal interactions are included in the model, there 
will be mixing between the two charged scalars.
 The $\delta a_e$ will receive additional contributions that is enhanced by 
${m_i}/{m_e}$ and it leads to a much more stringent constraint on $\kappa_1$ or $\kappa_2$. 
In this case, an $\mathcal{O}(1)$ magnitude of $\kappa_1$ or $\kappa_2$ can only be obtained for 
very small or maximal mixing angles~\cite{Chao:2016lqd}.

\begin{figure}[t]
\centering
   \includegraphics[width=0.9\columnwidth]{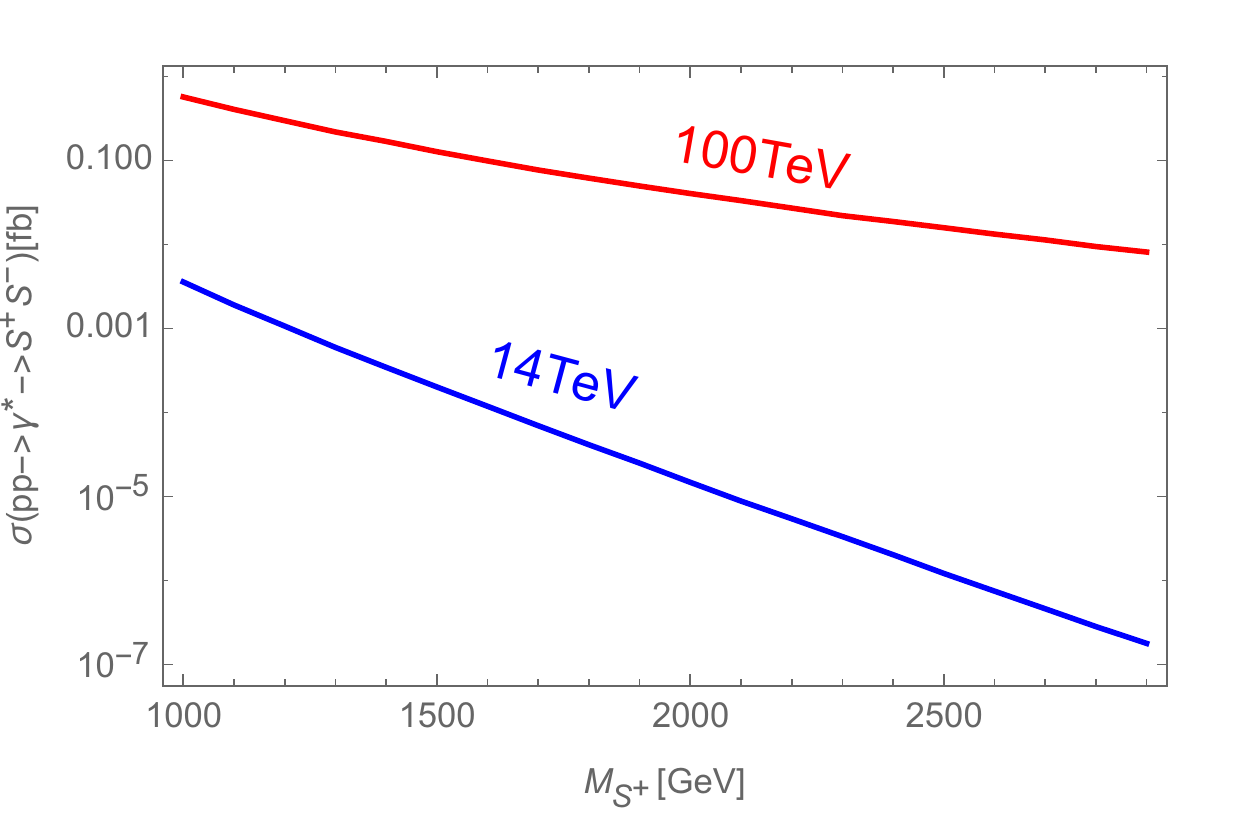}
 \caption{\label{fig:xsec}
   The production cross-section of a pair of charged scalar at 13 TeV LHC and 100 TeV proton-proton collider.
 }
 \end{figure}

The collider constraints mainly come from the search of a charged scalar. 
In LHC, the search for the production of a singly charged scalar is always associated with the production of a top quark which is motivated by the two Higgs doublet models.
 However, in our model the charged scalars only couple to leptons, so the search for production of a singly charged scalar puts no constraint on our model. 
 The other possibility is to search for the production of a pair of charged scalars which decay to 
 two opposite sign electrons and missing transverse energy. 
 This kind of signature is the same as the search for the pair production of the sleptons, which susbquently decay to SM leptons and neutralino. 
Current constraint~\cite{Khachatryan:2014qwa,Aad:2014vma} on the mass of slepton is around 200 GeV, while in our model, we focus on a charged scalar much heavier. So this constraint can be satisified. 
 As an illustration, we calculate the production cross-section of a pair of charged scalars at 13 TeV at LHC 
 and 100~TeV at $pp$ collider for the mass of charged scalar ranging from 1 TeV to 3 TeV using Madgraph~\cite{Alwall:2014hca} and show the plot in Fig.~\ref{fig:xsec}. 
One can observe that even if with 100 fb$^{-1}$ luminosity, the expected number of event at the 13 TeV LHC is less than $1$ for the charge scalar mass larger than 1.5 TeV.

\section{Relic density and Direct Detection}

The relic density of the DM has been precisely measured by Planck from a fit of 
the CMB anisotropies predicted by the $\Lambda \text{CDM}$ model with the observed data, 
giving $\Omega_{\text{c}} h^2 = 0.1199\pm 0.0022$~\cite{Ade:2015xua}.
In the freezing-out picture, the relic density of the DM is determined from the Boltzmann equation 
of the DM number density $n$~\cite{Gondolo:1990dk},
\begin{eqnarray}
\dot{n} + 3 Hn =- \langle \sigma v_{\rm M\slashed{o}ller} \rangle ( n^2 -n_{\rm EQ}^2 ) \; ,
\end{eqnarray}  
where $H $ is the Hubble constant, $\sigma v_{\rm M\slashed{o}ller}$ is the total annihilation cross section multiplied by  
the M$\slashed{\rm o}$ller velocity: $v_{\rm M\slashed{o}ller}=(|v_1 -v_2 |^2 -|v_1 \times v_2 |^2 )^{1/2}$, the brackets denote 
thermal average and $n_{\rm EQ}$ is the number density in thermal equilibrium. 
For both models, the results of $\langle\sigma v\rangle$ with the non-relativistic expansion $\langle \sigma v \rangle =a + b \langle v^2 \rangle$ in the lab frame can be written in the same form~\cite{Chao:2016lqd}:
\begin{eqnarray}
\langle \sigma v\rangle &=& \sum_I \zeta_i^2  \Big[ {m_\psi^2 \over 32 \pi (m_\psi^2 + {m}_{i}^2)^2 }  \nonumber \\
&& + \langle v^2 \rangle {m_\psi^2 (-7 m_\psi^4 -18 m_\psi^2 {m}_i^2 +{m}_i^4 )\over 384 \pi (m_\psi^2 + {m}_i^2)^4}\Big] .
\label{thermalaverage}
\end{eqnarray}
With this result, the relic abundance of the DM is thus given in terms of the dimensionless fraction of 
the critical energy density of the universe:
\begin{eqnarray}
  \Omega_c h^2 \approx 2 \times{1.07\times 10^9 \over M_{\text{pl}}} {x_F \over \sqrt{g_*} } {1 \over a + 3 b/x_F} ,
\end{eqnarray}
where $M_{\rm  pl}$ is the Planck mass, $x_F=m_{\rm DM}/T_F$ with
$T_F$ the freezing  out temperature of the DM and $g_*$ is the degrees of freedom at $T_F$.  
As mentioned earlier, for model I both channels, namely $ \psi \bar \psi \to e^+e^-$ and $\psi \bar \psi \to \nu_e \bar{\nu}_e$, contribute and the summation 
is over ${\Phi}^+$, $\rho$ and $\eta$, where $\zeta_{{\Phi}^+} = \kappa_1^2$ and
$\zeta_{\rho} = \zeta_{\eta} = \sqrt{2}\kappa_1^2$. 
On the other hand, for model II, 
only the channel $ \psi \bar \psi \to e^+ e^-$ is present, so that the factor
$\zeta_{{S}^+} = \kappa_2^2$. 
Note when there is mixing between the charged scalars, there is an 
additional contribution with the multiplication factor $(\kappa_1^2-\kappa_2^2) s_{2\theta}^2$ where 
$\theta$ is the mixing angle between the charged scalars~\cite{Chao:2016lqd}.

\begin{figure*}[t]
\centering
   \includegraphics[width=0.4\textwidth]{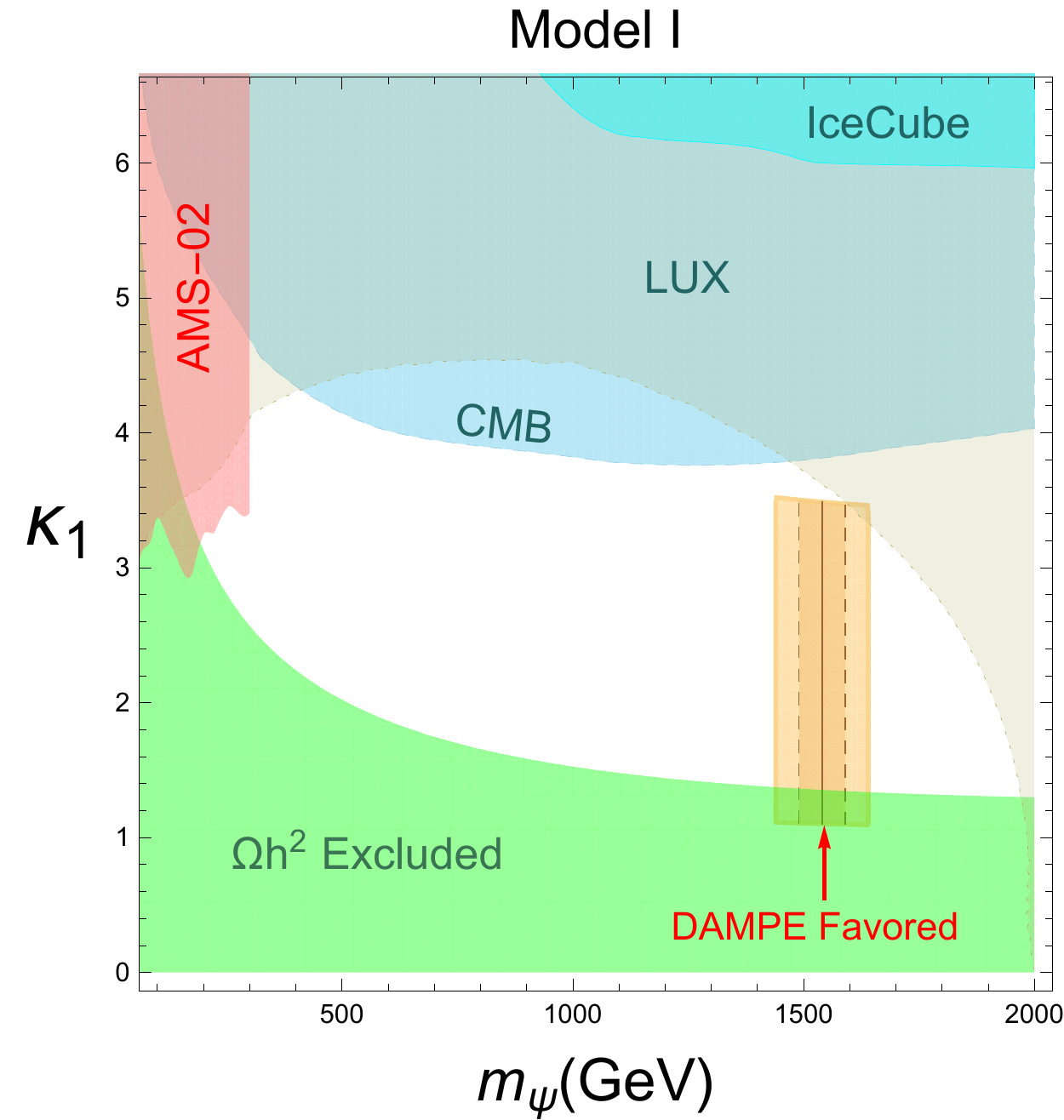}
   \quad\quad
   \includegraphics[width=0.4\textwidth]{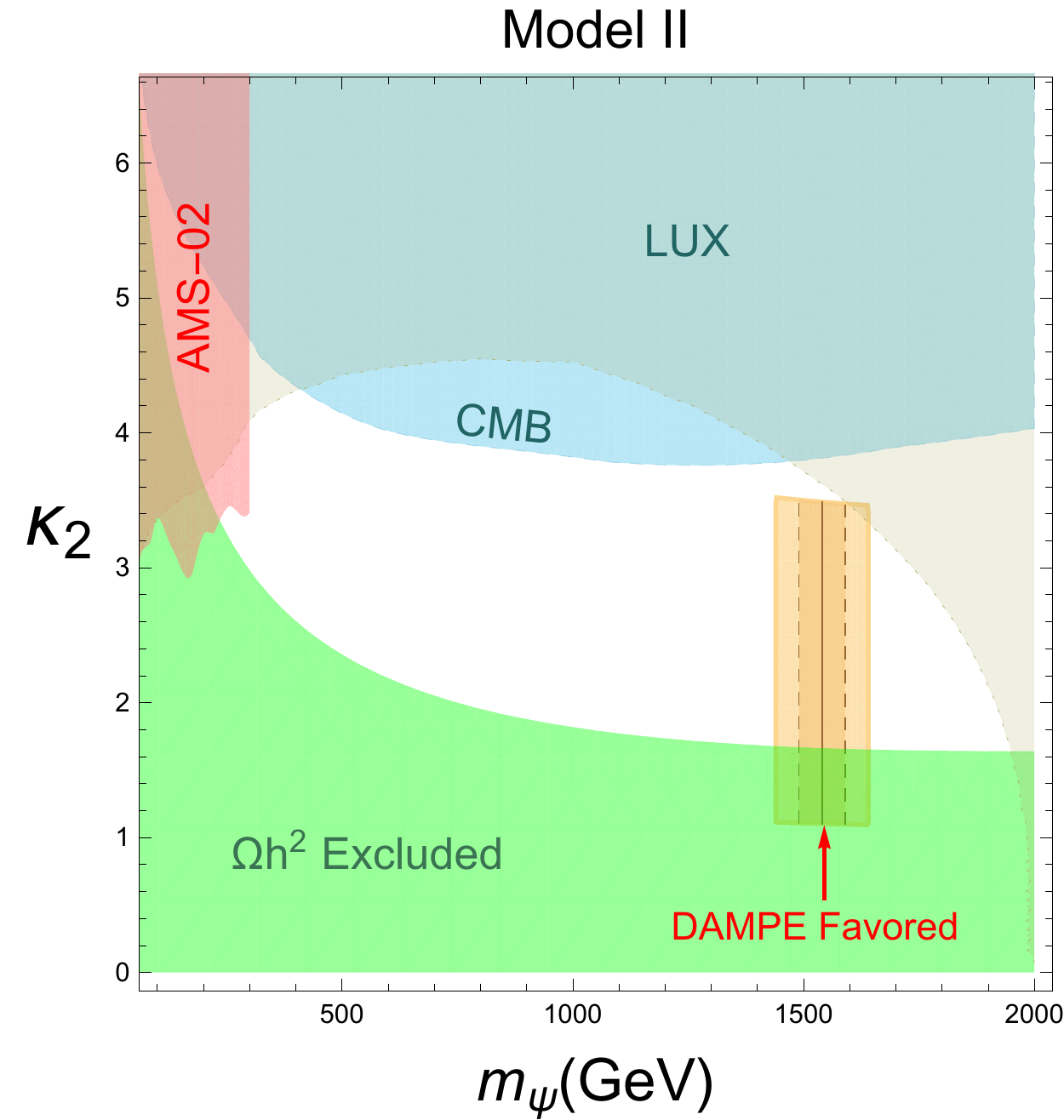}
 \caption{\label{fig:combined}
   Combined constraints on the portal couplings $\kappa_1$ and $\kappa_2$ for model I(left) and model
   II(right) respectively. In each case, the charged scalar mass is taken to be $2\text{TeV}$ and further in model I 
the neutral scalars both have mass $3\text{TeV}$. The excluded regions by CMB, LUX and AMS-02 are shown 
as color-shaded regions. The green region gives a larger relic density than the Planck result and is therefore
excluded.  The brown region, labelled as ``DAMPE Favored'', denotes the possible DM explaination of the 
$\approx 1.5\text{TeV}$ excess with $10^{-26}\lesssim \langle \sigma v \rangle_{e^+ e^-}/  \text{cm}^3 \text{s}^{-1}\lesssim 10^{-24}$.
 }
 \end{figure*}

For model I, by fixing $m_{\Phi^+} = 2\text{TeV}$ and $m_{\rho,\eta}=3\text{TeV}$, we show in the 
$\kappa_1 - m_{\psi}$ plane the green region that gives a larger relic density than the result given
by Planck in Fig.~\ref{fig:combined}. This region is denoted as ``$\Omega h^2$ \text{Excluded}''.
For model II, we fix $m_{S^+} = 2\text{TeV}$ and show the same region in the $\kappa_2 - m_{\psi}$ plane 
in the right panel of Fig.~\ref{fig:combined}. 
Note that the upper bound of this region which corresponds to the parameter choice giving the Planck value, 
is a bit higher than that in the left panel, because DM only annihilates into $e^+e^-$ in this case, while in model I  there is an additional annihilation channel $\psi \bar \psi \to \nu_e \bar{\nu}_e$.

Direct detection experiments in deep underground laboratories measure the nuclear recoils of the nuclei 
from elastic scattering of the DM particle with nucleus. 
Null search signal in such experiment puts upper limit on the 
spin-dependent and spin-independent cross sections of nucleon scattering off the DM as a function of the DM 
mass. 
Especially, recent years it has witnessed great improvements of such limits made by the PandaX-II~\cite{Cui:2017nnn}, Xenon1T~\cite{Aprile:2017iyp} and LUX~\cite{Akerib:2016vxi} experiments.
In the present context, the lepton flavored DM interacts with nucleons as the DM possesses non-trivial electromagnetic form factors induced 
from diagrams shown in Fig.~\ref{ddloopgamma}.

%
%


These interactions in Eq.~\ref{eq:dm-photon} translate, at the nucleon level, 
into the following effective Lagrangian~\cite{Chang:2014tea,Hamze:2014wca,Ibarra:2015fqa},
\begin{eqnarray}
&& f_r\bar \psi \gamma^\mu \psi \bar N \gamma_\mu N +f_h \bar \psi \psi \bar N N + f_m^1\bar \psi i\sigma^{\mu\nu} \psi { q_\nu \over q^2 } \bar N K_\mu N  \nonumber\\
&& +f_m^2 \bar \psi i\sigma^{\alpha\mu} \psi { q_\alpha q_\beta \over q^2 } \bar N i\sigma^{\beta \mu} N , 
  \label{effectnucl}
\end{eqnarray}
where $q^{\mu}$ is the momentum transfer from nucleon to DM, $K^{\mu}$ is the summation of momenta of 
incoming and outgoing nucleons. The Wilson coefficients above are given by 
\begin{eqnarray}
  &&  f_r^N = e Q_N b_\psi \; , \hspace{0.1cm} f_h^N = f_\psi^h {m_N \over m_h^2 v}\left( \sum_{q=u,d,s}  f_{T_q}^N + {2\over 9} f_{TG}^N\right)\; ,  \nonumber\\
  && f_m^1 = \frac{e Q_N \mu_\psi}{2m_N}\; ,  \hspace{0.1cm} f_m^2 = -\frac{e \tilde{\mu}_N \mu_\psi}{2m_N} ,\nonumber 
\end{eqnarray}
where $Q_N$ is the charge of the nucleon, $\mu_\psi$ and $b_\psi$ are the magnetic moment and charge radius of the DM 
respectively, Moreover $\tilde \mu_N $ is the nucleon magnetic moment: $\tilde{\mu}_p \approx 2.80$ and $\tilde{\mu}_n \approx -1.91$. 
Finally $f_{\psi}^h$ is the effective DM-Higgs coupling and is generally negligible relative to other 
contributions. 

Quite different from the standard spin-independent and spin-dependent description of the DM-nucleon 
interactions, above electromagnetic form factors of the DM induces distinctivly new nuclear responses. 
This is described in the EFT framework~\cite{Fan:2010gt,Fitzpatrick:2012ix,Anand:2013yka} and is implemented 
in the public code~\cite{DelNobile:2013sia} which we use in our numerical analysis. The resulting excluded 
region by LUX~\cite{Akerib:2016vxi} is shown in Fig.~\ref{fig:combined} for both model I and model II.
The difference between these two is not so noticeable since it is the magnetic moment($\mu_{\psi}$) and 
charge radius($b_{\psi}$) terms in Eq.~\ref{eq:dmresults} that make the dominant contribution, which have 
the same dependence on the charged scalar mass and coupling.

\section{Indirect Detections}
DM detectors in space search for anomalous fluxes of cosmic rays, positrons or antiprotons above the 
astrophysical background and infer the existence of the DM particle indirectly. A review of the current 
status of indirect detections is given in Ref.~\cite{Slatyer:2017sev}. 
In particular the 
AMS-02 in the $e^+ e^-$ channel have put constraint on $\langle \sigma v \rangle_{e^+ e^-}$ for DM up to $300\text{GeV}$~\cite{Bergstrom:2013jra}. 
This is shown as the 
pink region in Fig.~\ref{fig:combined}.
On the other hand, the annihilation of DM into charged final states such as $e^+ e^-$, in the dark ages after
the last scattering of photons, changes the standard reionization of the universe and affects the observed 
CMB anisotropies. 
This is used to help Planck put an upper limit on 
$f_{\text{eff}}\langle \sigma v \rangle$~\cite{Ade:2015xua}, 
where $f_{\text{eff}}$ is an efficiency factor of 
ionization~\cite{Finkbeiner:2011dx,Slatyer:2015jla,Slatyer:2015kla}. 
The resulting exclusion regions from 
$\langle \sigma v \rangle_{e^+ e^-}$ by CMB is shown as light-blue regions in Fig.~\ref{fig:combined}. 
Furthermore, IceCube have put an upper limit on $\langle \sigma v \rangle_{\nu \nu}$~\cite{Aartsen:2016pfc}, which constrains 
model I. 
This constraint is shown as the cyan region in the left panel of Fig.~\ref{fig:combined} for model I.
For model II, it is free from this limit. 
It can be seen from Fig.~\ref{fig:combined} that a significant 
proportion of the region in $\kappa_1 - m_{\psi}$ for model I and $\kappa_2 - m_{\psi}$ for model II are 
free from above indirect detection constraint as well as the LUX exclusion limit. 
Therefore both the electron
flavored DM scenarios can have the right relic density over a wide range of DM mass.

Finally, we comment on the recent DAMPE result. 
From a fit to the DAMPE data points~\cite{Chao:2017yjg}, it is found that for $\psi \psi \rightarrow e^+ e^-$, the DM parameter space is favored for a mass around 1.5 $\text{TeV}$ with $10^{-26}  \text{cm}^3 \text{s}^{-1} \lesssim \langle \sigma v \rangle_{e^+ e^-} \lesssim 10^{-24} \text{cm}^3 \text{s}^{-1}$. 
This requirement can be translated to the constraints of the model parameters, such as coupling $\kappa_1$ or $\kappa_2$. For $1400\text{GeV}< m_{\psi} < 1600\text{GeV}$, this appears as a region and is shown in Fig.~\ref{fig:combined}
as the brown region for model I (left) and model II (right) respectively. For this ``DAMPLE Favored'' region, 
the vertical solid line in the center denotes $m_{\psi}=1.5\text{TeV}$ while the two dashed lines
are for $1.5\text{TeV} \pm 50\text{GeV}$. From this figure, it is clear that this ``DAMPE Favored'' region 
overlaps the relic density contour at $\kappa_1 \approx 1.5$ for model I and $\kappa_2 \approx 1.7$ 
for model II, both of which are allowed by other indirect detection and direct detection constraints. 
We conclude that both model I and model II of the electron flavored DM scenarios discussed in this work 
are capable to explain the DAMPE excess through DM annihilation. Future accumulation of the data points by DAMPE will give us more information on this region.

\section{Summary}
In this work, we studied the electron-flavored DM which is realized in two models: model I and model II with different portal interactions with left-handed and right-handed electrons. We explored in detail the phenomenology of DM annihilations, DM direct detections and indirect detections as well as constraint from electron magnetic dipole moment. We found that both of the electron flavored DM scenarios considered here have a wide range of parameter space that is compatible with the current data. The most recent DAMPE excess could be interpreted as the DM annihilations in both scenarios with the DM mass around 1.5 TeV and annihilation cross section $10^{-26}  \text{cm}^3 \text{s}^{-1} \lesssim \langle \sigma v \rangle_{e^+ e^-} \lesssim 10^{-24} \text{cm}^3 \text{s}^{-1}$.

\section{Acknowledgements}
We thank Qiang Yuan and Yong Du for helpful discussions.
W.C. is supported by the National Natural Science 
Foundation of China under grant No.11775025 and the Fundamental Research Funds 
for the Central Universities. 
J.S. is supported by the NSFC under grant No.11647601, No.11690022 and No.11675243 and also supported by the Strategic Priority Research Program of the Chinese Academy of Sciences under grant No.XDB21010200 and No.XDB23030100.


\end{document}